\documentclass[12pt]{iopart}

\newtheorem{lemma}{Lemma}
\newtheorem{proposition}{Proposition}
\newtheorem{theorem}{Theorem}

\newtheorem{definition}{Definition}
\newcommand{\be}{\begin{equation}}
\newcommand{\ee}{\end{equation}}

\newcommand{\ci}{\mathop{\textrm{i}}}

\begin{document}

\title[Bel radiative gravitational fields] {On the
Bel radiative gravitational fields}

\author{Joan Josep Ferrando$^1$\
and Juan Antonio S\'aez$^2$}

\address{$^1$\ Departament d'Astronomia i Astrof\'{\i}sica, Universitat
de Val\`encia, E-46100 Burjassot, Val\`encia, Spain.}

\address{$^2$\ Departament de Matem\`atiques per a l'Economia i l'Empresa,
Universitat de Val\`encia, E-46071 Val\`encia, Spain}

\ead{joan.ferrando@uv.es; juan.a.saez@uv.es}

\begin{abstract}
We analyze the concept of intrinsic radiative gravitational fields
defined by Bel and we show that the three radiative types, N, III
and II, correspond with the three following different physical
situations: {\it pure radiation}, {\it asymptotic pure radiation}
and {\it generic} (non pure, non asymptotic pure) {\it radiation}.
We introduce the concept of {\em observer at rest} with respect to
the gravitational field and that of {\em proper super-energy} of the
gravitational field and we show that, for non radiative fields, the
minimum value of the relative super-energy density is the proper
super-energy density, which is acquired by the observers at rest
with respect to the field. Several {\it super-energy inequalities}
are also examined.
\end{abstract}




\section{Introduction}
\label{sec-intro}

With the purpose of defining intrinsic states of gravitational
radiation Bel \cite{bel-1, bel-2, bel-3} introduced a rank 4 tensor
which plays an analogous role for gravitation to that played by the
Maxwell-Minkowski tensor for electromagnetism. In the vacuum case
this {\it super-energy Bel tensor} is divergence-free and it
coincides with the {\it super-energy Bel-Robinson tensor} $T$.

Using tensor $T$, Bel defined the relative {\it super-energy
density} and the {\it super-Poynting vector} associated with an
observer. Then, following the analogy with electromagnetism, the
intrinsic radiative gravitational fields are those for which the
Poynting vector does not vanish for any observer \cite{bel-1,bel-3}.
The Bel approach, based on super-energy concepts, leads to the same
gravitational fields as the Pirani \cite{pirani} one, which is based
on intrinsic geometric considerations (see \cite{bel-3, zak} for
these and other radiation criteria).

It is worth remarking that Bel super-energy quantities do not
represent gravitational energy. Nevertheless, the relationship
between super-energy and quasi-local gravitational energy has been
largely discussed \cite{horowitz} (a wide list of references on this
subject can be found in \cite{seno} and \cite{szabados}).

The interest of the Bel approach has recently been remarked by
Garc\'{\i}a-Parrado \cite{garcia-parrado}, who introduces new
relative super-energy quantities and writes the full set of
equations for these super-energy quantities. This study leads
naturally to a concept of intrinsic radiation which is less
restrictive than Bel's \cite{garcia-parrado}. We will analyze
Garc\'{\i}a-Parrado's proposal in a forthcoming work \cite{fsRS-b}
where we will also give an intrinsic characterization of the new
radiative classes.

It is worth remarking that Bel and Garc\'{\i}a-Parrado definitions
are local, and a gravitational field radiative at a point of the
space-time could be non radiative at another point. In this work we
give several definitions and results that are also local because
they are based on algebraic considerations.

Here we revisit Bel's ideas in depth. The analogy with
electromagnetism helps us understand the already known concepts and
the new ones we are introducing. For this reason we devote section
\ref{sec-2} to summarize several known results on the
electromagnetic field. We present them in a way that facilitates
their extension to the gravitational field.

In section \ref{sec-3} we introduce the notation used in this work
for the super-energy quantities defined by Bel and
Garc\'{\i}a-Parrado. In section \ref{sec-4} we give a set of
super-energy inequalities which extend the previously known ones.
These inequalities are used to prove our essential results.

In section \ref{sec-5} we define the {\it pure radiative}, {\it
asymptotic pure radiative} and {\it generic radiative} gravitational
fields, and we show that these three different physical situations
correspond to the three Bel radiative cases, namely, the Petrov-Bel
types N, III and II, respectively. The asymptotic behavior of every
radiative type is also analyzed.

Section \ref{sec-6} is devoted to studying non radiative fields.
Extending to the gravitational field the concepts of observer at
rest and of proper energy density introduced for the electromagnetic
field by Coll \cite{tolo-em}, we define the concepts of {\it
observer at rest} with respect to a gravitational field and of {\it
proper super-energy density}. We show that, for a non radiative
field, the minimum value of the super-energy density is the proper
one and it is attained for the observers at rest. The proper
super-energy density is also analyzed for radiative fields.

Finally, we present three appendices. The first one explains some
notation. The second one summarizes the canonical forms of the
Bel-Robinson tensor, while the third one gives the accurate proof of
the main theorem stated in section \ref{sec-4}.

\section{Intrinsic radiative electromagnetic fields}
\label{sec-2}

The concept of radiative electromagnetic fields is well known.
Electromagnetic radiative states are modeled by null electromagnetic
fields. In this section we revisit this topic summarizing several
known definitions and properties. We also point out that some
concepts, like {\it field of pure radiation} and {\it field of
intrinsic radiation} which are concurrent in the electromagnetic
case, must be considered as conceptually different. Then, these
differences could be important when analyzing super-energy radiative
gravitational fields.

\subsection{The Faraday and Maxwell-Minkowski tensors. Relative
formulation} \label{subsec-2-a}

We shall note $g$ the space-time metric with signature convention
$(-,\, +, \, +,\,+)$. The electromagnetic field is modeled with the
Faraday 2-form $F$. The electromagnetic energy-momentum
(Maxwell-Minkowski) tensor $M$ is given in terms of $F$ by
$M=-\frac12[F^2 + *F^2]$, where $*$ stands for the Hodge dual
operator.

For any observer (unitary time-like vector) $u$, the relative {\it
electric} and {\it magnetic} fields are given by $e= F(u)$ and
$b=*F(u)$, respectively. And the relative {\it energy density}
$\rho$, {\it Poynting vector} $s_{\perp}$ and {\it electromagnetic
stress tensor} $M_{\perp}$ are given by:
$$
\rho = M(u,u) = \frac12(e^2 + b^2)  ,\ \ \ s_{\perp} = -M(u)_{\perp}
=
*(u \wedge e \wedge b)  , \ \ \  M_{\perp} = \rho \gamma
- e \otimes e - b \otimes b  ,
$$
where, for a tensor $A$, $A_{\perp}$ denotes the orthogonal
projection defined by the projector $\gamma = u \otimes u + g$. In
terms of these relative energetic variables the Maxwell-Minkowski
tensor $M$ takes the expression:
\begin{equation}
M = M_{\perp} + s_{\perp} \stackrel{\sim}{\otimes} u + \rho \, u
\otimes u \, .
\end{equation}
where, for two vectors $a, b$, $a \stackrel{\sim}{\otimes} b = a
\otimes b + b \otimes a$.

When $F$ is a Maxwell field, $dF=0, \ \nabla \cdot F=0$, $M$
satisfies the conservation equation $\nabla \cdot M=0$. For any
observer, the time-like component of this equation shows that the
relative vector $s_{\perp}$ is, indeed, the flux of the relative
scalar $\rho$.

\subsection{Algebraic restrictions and Pleba\'{n}ski energy conditions}
\label{subsec-2-b}

The electromagnetic energy tensor $M$ satisfies the {\it algebraic
Rainich conditions} \cite{rainich}:
\begin{equation}
\tr M = 0 \, , \qquad M^2 = \chi^2 g \, , \quad  \chi \equiv \frac12
\sqrt{\tr M^2} \, , \qquad M(x,x) \geq 0 \, ,
\end{equation}
where $x$ is any given time-like vector. The last one states that
the {\it weak energy condition} holds, that is, the energy density
is non-negative for any observer.

A significant property is that the above Rainich conditions imply
$M^2(x,x) \leq 0$ for an arbitrary time-like vector $x$. This means
that the energy-momentum density, $s = - M(u) = s_{\perp} + \rho u$,
is a causal vector for any observer $u$. This is a physical
requirement which expresses that the amount of radiating energy is a
part of the total energy:
\begin{equation} \label{chi-s-ro}
\rho^2 - s_{\perp}^2 = -s^2 = \chi^2 \geq 0  \, .
\end{equation}
Note that the scalar $\sqrt{-s^2}$ built with the relative magnitud
$s$ is the electromagnetic invariant $\chi$, which has been called
the {\it proper energy density} of the electromagnetic field
\cite{tolo-em}.

It is worth remarking that Pleba\'{n}ski energy conditions ($M(x,x)
\geq 0, M^2(x,x) \leq 0$) \cite{plebanski} restrict any energy
tensor $M$ to be (real) type I or type II with additional
constraints on its eigenvalues. Then the invariant $\tr M^2$ is
non-negative. Thus, generically, the Pleba\'{n}ski energy conditions
for an energy tensor $M$ state, $\rho \geq 0$, $s^2\leq 0$, $\tr M^2
\geq 0$. In the case of the electromagnetic field these quantities
are bounded by the proper energy $\chi$.
\begin{proposition} \label{prop-ec-e}
{\bf (energy conditions)} Let $M$ be the Maxwell-Minkowski energy
tensor and for any observer $u$ let us define the relative
space-time quantities:
\begin{equation} \label{M-s-rho}
s = -M(u)\,  ,\qquad  \rho = M(u,u) \, .
\end{equation}
Then, the following energy conditions hold:
\begin{equation} \label{s-ec}
\tr M^2 \equiv 4 \chi^2 \geq 0 \,  , \qquad s^2 = - \chi^2 \leq 0 \,
, \qquad  \rho \geq \chi \geq 0 \, .
\end{equation}
\end{proposition}

\subsection{Intrinsic radiative states. Null fields}
\label{subsec-2-c}

The energy density $\rho$ and the stress tensor $M_{\perp}$ are
related by $\tr M_{\perp} = \rho$, and they vanish only when $F$
vanishes. Thus, any of these relative quantities enable an observer
to know if an electromagnetic field is present or not, and they give
a measure of the intensity of this field. Nevertheless, the Poynting
vector $s_{\perp}$ relative to an observer can vanish for non zero
electromagnetic fields. This fact enables us to distinguish a
special class of fields:

\begin{definition} \label{def-irs-e}
An energy tensor $M$ represents a state of {\em intrinsic radiation}
(at a point) when the Poynting vector $s_{\perp}$ does not vanish
for any observer.
\end{definition}

\begin{proposition} \label{prop-irs-e}
The intrinsic radiative electromagnetic fields are the {\em null fields}
which are characterized by one of the following equivalent
conditions:

(i) The invariant (proper energy density) $\chi \equiv \frac12
\sqrt{\tr M^2}$ vanishes.

(ii) The electric and magnetic fields are orthogonal, $(e, h) = 0$,
and equimodular, $e^2 = h^2$.
\end{proposition}

For a better understanding of the intrinsic radiation states for
both the electromagnetic and gravitational fields we give the
following definition.

\begin{definition} \label{def-prs-e}
An energy tensor $M$ represents a state of {\em pure radiation} (at
a point) when the whole energy density is radiated as Poynting
energy, $\rho = |s_{\perp}|$.

An energy tensor $M$ represents a state of {\em asymptotic pure
radiation} (at a point) when $\rho \not= |s_{\perp}|$ and for any
positive real number $\epsilon$ one can find an observer for which
the non radiated energy $\rho - |s_{\perp}|$ is smaller than
$\epsilon$.
\end{definition}
As a consequence of proposition \ref{prop-irs-e} we have:

\begin{proposition} \label{prop-prs-e}
All the intrinsic radiative electromagnetic fields are of pure
radiation.
\end{proposition}

The concept of asymptotic pure radiation does not give a new class
in the electromagnetic case because $\rho^2 - s_{\perp}^2$ is an
invariant. Nevertheless, we will see in this work that the
definitions given above distinguish three different super-energy
radiative gravitational fields.

The energy tensor of a null electromagnetic field takes the
expression $M= \ell \otimes \ell$, where $\ell$ is the (like-light)
{\it fundamental vector} of the null field. For any observer, the
energy-momentum density $s$ points the fundamental direction. More
precisely, for any observer $u$, the fundamental vector takes the
expression $\ell = \sqrt{\rho} (u + e)$, where $\rho$ is the
relative energy density and $e$ is the unit vector pointing out the
spatial direction of propagation of radiation, $\ell_{\perp} =
\sqrt{\rho}\, e \propto s_{\perp}$. Then, the observers $\tilde{u}$
traveling (with respect to $u$) in the direction (respectively,
opposite direction) of $s_{\perp}$ are $\tilde{u} = \cosh \varphi\,
u + \sinh \varphi\, e$, with $\varphi > 0$ (respectively, $\varphi <
0$), and the relative energy density is $\tilde{\rho} = \rho \, e^{-
2\varphi}$. Consequently, we obtain:

\begin{proposition} \label{prop-fv-e}
The fundamental vector $\ell$ of a null electromagnetic field
determines the spatial direction of propagation of radiation, that
is, for any observer, $\ell_{\perp} \propto s_{\perp}$.

For a family of observers having each a spatial velocity at a point
tangent and parallel (opposite) to $\ell_{\perp}$, the energy
density measured by an observer at the same point decreases
(increases) and tends to zero (infinity) as its spatial velocity
increase and approaches the speed of light. A similar conclusion
holds for the radiated energy.

\end{proposition}

\subsection{Non intrinsic radiative states. Observer at rest and proper
energy density} \label{subsec-2-d}

For a non intrinsic radiative electromagnetic field at least one
observer exists that sees a vanishing relative Poynting vector.

\begin{proposition} \label{prop-nirs-e}
 The non intrinsic radiative electromagnetic fields are the {\em
non null fields} which are characterized by having a non vanishing
proper energy density, $\chi \equiv \frac12 \sqrt{\tr M^2}\not=0$.
\end{proposition}
The following definition naturally arises \cite{tolo-em}:

\begin{definition} \label{def-or-e}
The observers that see the proper energy density $\chi$ as their
energy density, for which the Poynting vector vanishes, are said
{\em observers at rest} with respect to the electromagnetic field.
\end{definition}

The energy tensor of a non null electromagnetic field takes the
expression $M= -\chi(v - h)$, where $v$ (respectively, $h$) is the
projector on the time-like (respectively, space-like) {\it principal
plane}. Then, we have:

\begin{proposition} \label{prop-or-e}
 The observers at rest with respect to a non null electromagnetic field
are those lying on the time-like principal plane.
\end{proposition}

On the other hand, from expression (\ref{chi-s-ro}) we obtain:

\begin{proposition} \label{prop-min-e}
 For a non null electromagnetic field the minimum value of the relative
energy density is the proper energy density $\chi$, which is
acquired by the observers at rest with respect to the field.
\end{proposition}

\section{The Weyl and Bel-Robinson tensors. Relative formulation}
\label{sec-3}

In vacuum, the intrinsic properties of a gravitational field depend
on the Weyl tensor $W$. Then, the Bel tensor coincides with the
Bel-Robinson tensor given in terms of $W$ as \cite{bel-1, bel-2,
bel-3}:
\begin{equation} \label{BR-1}
{T_{\alpha  \mu \beta \nu}} = \frac14 \left({{{
W_{\alpha}}^{\rho}}_{\beta}}^{ \sigma} W_{\mu \rho \nu \sigma } +
{{{* W_{\alpha}}^{\rho}}_{\beta}}^{ \sigma} *\! W_{\mu \rho \nu
\sigma }\right) \, ,
\end{equation}

For any observer $u$, the relative {\it electric} and {\it magnetic}
Weyl fields are given by $E= W(u;u)$ and $H= *W(u;u)$, respectively.
The relative {\it super-energy density} $\tau$, {\it
super-Poynting (energy flux) vector} $q_{\perp}$, {\it super-stress
tensor} $t_{\perp}$, {\it stress flux tensor} $Q_{\perp}$ and {\it
stress-stress tensor} $T_{\perp}$ are given by:
$$
\tau = T(u,u,u,u)\,  ,\ \ \ q_{\perp} = -T(u,u,u)_{\perp} \, ,\ \ \
t_{\perp} = T(u,u)_{\perp}\,  ,\ \ \ Q_{\perp} = -T(u)_{\perp} \, ,
\ \ \ T_{\perp} \, .
$$
Bel introduced $\tau$ and $q_{\perp}$ years ago \cite{bel-1, bel-3},
and recently Garc\'{\i}a-Parrado \cite{garcia-parrado} has
considered $t_{\perp}$, $Q_{\perp}$ and $T_{\perp}$ giving their
expressions in terms of the electric and magnetic Weyl fields.

In terms of these relative super-energetic variables the Bel-Robinson
tensor $T$ takes the expression \cite{garcia-parrado}:
\begin{equation}
\hspace{-2.5cm} T = T_{\perp} + \, 4\; ^{\rm s}\!\{Q_{\perp}\!
\otimes u\} + \, 6\; ^{\rm s}\!\{t_{\perp}\! \otimes u \otimes u\} +
\, 4\; ^{\rm s}\!\{q_{\perp}\! \otimes u \otimes u \otimes u\} +
\tau \, u \otimes u \otimes u \otimes u \, ,
\end{equation}
where $^{\rm s}\!A$ denotes the symmetrization of a tensor $A$.

In vacuum, the Bianchi identities imply that $T$ satisfies the
conservation equation $\nabla \cdot T=0$. For any observer, this
equation shows that the relative quantities $q_{\perp}$ and
$Q_{\perp}$ play the role of fluxes of the relative quantities
$\tau$ and $t_{\perp}$, respectively \cite{garcia-parrado}.

\section{Algebraic restrictions and super-energy inequalities}
\label{sec-4}

Elsewhere \cite{fsBR-1,fsBR-2} we have studied the Bel-Robinson
tensor $T$ as an endomorphism on the 9-dimensional space of the
traceless symmetric tensors. Its nine eigenvalues $\{ t_k, \tau_k ,
\bar{\tau}_k \} $ depend on the three complex Weyl eigenvalues $\{
\rho_k \}$ as
\begin{equation} \label{valprobel}
 t_k = | \rho_k |^2 ; \qquad \tau_k = \rho_i \bar{\rho}_j , \quad (ijk)
\equiv \ even \ permutation\ of \ (123) \, .
 \end{equation}
We have also intrinsically characterized the algebraic classes of
$T$ \cite{fsBR-1}, and we have given their Segr\`e type and their
canonical form \cite{fsBR-2}. The part of these results needed in
this work are summarized in \ref{B-Bel-Robinson}.

Bergqvist and Lankinen \cite{bergqvist-lan} obtained the algebraic
constraints on the Bel-Robinson tensor playing a similar role to
that played by the Rainich conditions for the electromagnetic energy
tensor:
\begin{equation} \label{ber-lan-condition}
\tr T = 0 \, , \qquad T \cdot T = {\cal B}(T^2) \, , \qquad
T(x,x,x,x) \geq 0  \, ,
\end{equation}
where $x$ is any given time-like vector. The last one implies that
the {\it weak super-energy condition} holds, that is, the
super-energy density is non-negative for an arbitrary observer. The
second one states that the six-order tensor $(T \cdot T)_{\alpha
\beta \gamma \lambda \mu \nu} = T_{\alpha \beta \gamma \sigma}
T^{\sigma}_{\ \lambda \mu \nu}$ depends on the four-order one
$(T^2)_{\alpha \beta \lambda \mu} = T_{\alpha \beta \gamma \sigma}
T^{\gamma \sigma}_{\ \ \lambda \mu}$. The explicit expression of the
linear operator ${\cal B}$ can be found in \cite{bergqvist-lan}. A
direct consequence of this constraint is the known relation:
\begin{equation} \label{trT2}
 \tr T^2 = \frac14 (T,T) g \, .
\end{equation}
Note that the quadratic scalar $(T,T) = T_{\alpha \beta \lambda \mu}
T^{\alpha \beta \lambda \mu}$ associated with $T$ is non-negative,
because of $64 (T,T)= (W,W)^2 + (W,*W)^2$ as a consequence of the
results in \cite{fsBR-1}.

The Bel-Robinson tensor $T$ also satisfies another super-energy
inequality, namely, $q = -T(u,u,u) = \tau u + q_{\perp}$ is a causal
vector for any observer $u$,
\begin{equation} \label{tau-q}
\tau^2 - q_{\perp}^2 = -q^2 = -(q,q) \geq 0  \, .
\end{equation}

It is worth mentioning that the super-energy inequalities $\tau \geq
0$ and $(q,q) \leq 0$ can be derived from a stronger condition which
satisfies the Bel-Robinson tensor: the generalized dominant energy
condition. A wide study about the dominant energy condition for
super-energy tensors and general tensors can be found in \cite{seno,
seno2}.

Condition $(q,q) \leq 0$ was shown by Bonilla and Senovilla
\cite{bonilla-sen} using the relative electric and magnetic Weyl
tensors, and it was recovered in \cite{bergqvist} exploiting the
spinorial formalism. Here, we present a stronger inequality working
with the Bel-Robinson tensor itself. Our tensorial proof is based in
the following main theorem which is shown in \ref{C-theor-main}.
\begin{theorem} \label{theo-main}
Let $T$ be the Bel-Robinson tensor and let us define its invariant
scalars:
\begin{equation} \label{alpha-xi}
\alpha \equiv \frac12 \sqrt{(T,T)} \, ,  \qquad  \xi  \equiv \frac14
\sum t_i \, ,
\end{equation}
where $t_i$ are the Bel-Robinson real eigenvalues. Then, for any
observer $u$,
\begin{equation} \label{T-alpha-xi}
T(u,u,u,u) \geq \xi \geq \frac12 \alpha \, ,\qquad T^2(u,u,u,u) \geq
\frac12 \alpha^2  \, .
\end{equation}
\end{theorem}

From here we can show stronger constraints for $\tau$ and $(q,q)$,
and also other super-energy inequalities that we collect in the
following statement.
\begin{theorem} \label{theo-ec-g}
{\bf (super-energy inequalities)} Let $T$ be the Bel-Robinson tensor
and for any observer $u$ let us define the relative space-time
quantities:
\begin{equation} \label{T-Q-t-q-tau}
Q = -T(u)\,  ,\quad t = T(u,u) \,  ,\quad q = -T(u,u,u) \, ,\quad
 \tau = T(u,u,u,u)  \, .
\end{equation}
Then, the following super-energy inequalities hold:
\begin{equation} \label{ec-g}
\hspace{-2.5cm} (T,T) \equiv 4 \alpha^2 \geq 0  ,  \ \ (Q,Q) = -
\alpha^2 \leq 0  ,  \ \  (t,t)  \geq \frac12 \alpha^2 ,  \ \ (q,q)
\leq -\frac14 \alpha^2 ,  \ \  \tau  \geq \frac12 \alpha \geq 0 .
\end{equation}
\end{theorem}
The first condition in (\ref{ec-g}) is the definition of the scalar
$\alpha$. The second one follows from (\ref{trT2}). The third and
the fifth ones come from (\ref{T-alpha-xi}). Finally, the fourth one
is a consequence of the following conditions:
\begin{equation} \label{t-q-tau}
3 (t_{\perp}, t_{\perp})- 2 (q_{\perp}, q_{\perp}) - \tau^2 =
\frac12 \alpha^2 \, , \quad (t_{\perp},t_{\perp})- 2
(q_{\perp},q_{\perp}) + \tau^2 \geq \frac12 \alpha^2 \, ,
\end{equation}
which come from the Bergqvist-Lankinen constraint
(\ref{ber-lan-condition}) and the third condition in (\ref{ec-g}),
respectively.

From the above restrictions (\ref{t-q-tau}) we also recover another
result by Bonilla and Senovilla used in the proof of the causal
propagation of gravity in vacuum \cite{bonilla-sen}:
\begin{proposition} \label{prop-Bon-Sen}
The amount of super-stress is bounded by the amount of super-energy,
$$
(t_{\perp}, t_{\perp}) \leq \tau^2 \, .
$$
\end{proposition}

\section{Intrinsic super-energy radiative gravitational fields}
\label{sec-5}

The super-energy density $\tau$, the super-stress tensor $t_{\perp}$
and the stress-stress tensor $T_{\perp}$ are related by $\tr
T_{\perp} = t_{\perp}$, $\tr t_{\perp} = \tau$, and they vanish only
when the Weyl tensor $W$ vanishes \cite{garcia-parrado}. Thus, any
of these relative quantities enable an observer to know if the
purely gravitational part of the field is present or not, and they
give a measure of the intensity of this field. Nevertheless, the
Poynting vector $q_{\perp}$ and the stress flux tensor $Q_{\perp}$
relative to an observer can vanish for a non zero Weyl tensor.

If we consider $\tau$ as a measure of the gravitational field, its
flux $q_{\perp}$ denotes the presence of gravitational radiation.
This is the point of view of Bel \cite{bel-1, bel-3}, who gave the
following definition.

\begin{definition} \label{def-bel-g}
{\bf (intrinsic gravitational radiation, Bel 1958)} In a vacuum
space-time there exists {\em intrinsic gravitational radiation} (at
a point) if the super-Poynting vector $q_{\perp}$ does not vanish
for any observer.
\end{definition}

But we can also consider $t_{\perp}$ as a measure of the
gravitational field. Then its flux $Q_{\perp}$ denotes the presence
of gravitational radiation. This fact has been pointed out by
Garc\'{\i}a-Parrado \cite{garcia-parrado}, who has given the
following definition.

\begin{definition} \label{def-gapa-g}
{\bf (intrinsic super-energy radiation, Garc\'{\i}a-Parrado 2008)}
In a vacuum space-time there exists {\em intrinsic super-energy
radiation} (at a point) if the stress flux tensor $Q_{\perp}$ does
not vanish for any observer.
\end{definition}

The criterion given by Bel leads to gravitational fields of
Petrov-Bel type N, III and II as modeling gravitational radiative
states \cite{bel-3} according with the Pirani's proposal
\cite{pirani}. On these and other radiation criteria see
\cite{bel-3,zak}.

The definition given by Garc\'{\i}a-Parrado is less restrictive than
the Bel one and it allows type I radiative gravitational fields
\cite{garcia-parrado}. This and other satisfactory properties show
the interest of this generalization which claims for a deeper study
undertaken elsewhere \cite{fsRS-b}. In this paper we focus on
analyzing Bel's radiative gravitational fields.

In his work on gravitational radiation Lichnerowicz \cite{lich}
considers type N gravitational fields as modeling {\it pure
radiation}, and Bonilla and Senovilla \cite{bonilla-sen} showed that
this case can be characterized by the condition $(q,q)=0$. Then, the
super-energy density $\tau$ equals the amount of radiating energy
$|q_{\perp}|$. Our definition \ref{def-prs-e} links this feature
with the Lichnerowicz terminology:

\begin{definition} \label{def-prs-q}
{\bf (pure gravitational radiation)} In a vacuum space-time there
exists {\em pure gravitational radiation} (at a point) when the
whole super-energy density is radiated as Poynting super-energy,
$\tau = |q_{\perp}|$.
\end{definition}

Note the definition we give is Bonilla-Senovilla's description of
the Lichnerowicz concept of pure radiation. Evidently, we have the
following intrinsic characterization \cite{bonilla-sen}:

\begin{proposition} \label{prop-N-charac}
The pure radiative states are the type N gravitational fields.
\end{proposition}

A type N space-time admits a quadruple null Debever direction $\ell$
which is named the {\it fundamental direction} of the gravitational
field. For any observer, the relative quantity $q$ points the
fundamental direction. Then, from the canonical expression of a type
N Bel-Robinson tensor, a similar reasoning to that we have done
before proposition \ref{prop-fv-e} leads to the following result.

\begin{proposition} \label{prop-N-l}
The fundamental direction $\ell$ of a type N gravitational field
determines the spatial direction of propagation of radiation, that
is, for any observer, $\ell_{\perp} \propto q_{\perp}$.

For a family of observers having each a spatial velocity at a point
tangent and parallel (opposite) to $\ell_{\perp}$, the super-energy
density measured by an observer at the same point decreases
(increases) and tends to zero (infinity) as its spatial velocity
increase and approaches the speed of light. A similar conclusion
holds for the radiated super-energy.

\end{proposition}

Now we analyze the non pure radiative states. The following
definition distinguishes a subclass:

\begin{definition} \label{def-aprs-q}
{\bf (asymptotic pure gravitational radiation)} In a vacuum
space-time there exists {\em asymptotic pure gravitational
radiation} (at a point) when $\tau \not= |q_{\perp}|$ and for any
positive real number $\epsilon$ one can find an observer for which
the non radiated energy $\tau - |q_{\perp}|$ is smaller than
$\epsilon$.
\end{definition}

The fourth super-energy condition given in expression (\ref{ec-g})
of theorem \ref{theo-ec-g} states $(q,q) \leq -\frac14 \alpha^2$. In
a type II space-time the Bel-Robinson real eigenvalues do not vanish
(see \ref{B-Bel-Robinson}) and consequently, $\alpha \not= 0$. Then,
non asymptotic pure radiation exists in this case.

Nevertheless, in a type III space-time $\alpha=0$. Moreover, a
triple null Debever direction $\ell$ (named the {\it fundamental
direction}) and a simple one $k$ exist. Then, taking into account
the canonical form of a type III Bel-Robinson tensor (see expression
(\ref{BR-canIII}) in \ref{B-Bel-Robinson}), a similar reasoning to
that we have done before proposition \ref{prop-fv-e} leads to the
following result.

\begin{proposition} \label{prop-III-charac-l}
The asymptotic pure radiative states are the type III gravitational
fields.

Let $\ell$ be the fundamental direction of a type III gravitational
field. For a family of observers having each a spatial velocity at a
point tangent and parallel (opposite) to $\ell_{\perp}$, the
super-energy density measured by an observer at the same point
decreases (increases) and tends to zero (infinity) as its spatial
velocity increase and approaches the speed of light. A similar
conclusion holds for the radiated super-energy.

Observers in the Debever plane $\{\ell, k\}$ are those for which
$\ell_{\perp}$ gives the spatial direction of propagation of
radiation, $\ell_{\perp} \propto q_{\perp}$.
\end{proposition}

Finally, the type II space-times model the generic radiative states.
Now a double null Debever direction (named the {\it fundamental
direction}) and two simple ones $k_1$ and $k_2$ exist. Then, taking
into account the canonical form of a type II Bel-Robinson tensor
(see \ref{B-Bel-Robinson}) we obtain (\ref{II-tau}) (see
\ref{C-theor-main}). Then, a similar reasoning to that we have done
before proposition \ref{prop-fv-e} leads to the following result.

\begin{proposition} \label{prop-II-charac-l}
The generic (non pure, non asymptotic pure) radiative states are the
type II gravitational fields.

Let $\ell$ be the fundamental direction of a type II gravitational
field. For a family of observers having each a spatial velocity at a
point tangent and parallel (opposite) to $\ell_{\perp}$, the
super-energy density measured by an observer at the same point
decreases (increases) and tends to a positive value (infinity) as
its spatial velocity increase and approaches the speed of light.
Meanwhile, radiated super-energy decreases (increases) and tends to
zero (infinity).

\end{proposition}

\section{Non radiative gravitational fields. Observer at rest and
proper super-energy density} \label{sec-6}

From the Bel's point of view, non radiative gravitational fields are
those for which at least an observer exists which sees a vanishing
relative super-Poynting vector. The following definition naturally
arises:

\begin{definition} \label{def-or-g}
The observers for which the super-Poynting vector vanishes, are said
{\em observers at rest} with respect to the gravitational field.
\end{definition}
Firstly, we have this immediate result \cite{bel-3}:

\begin{proposition} \label{prop-nirs-g}
The non intrinsic radiative gravitational fields are the Petrov-Bel
type I or D space-times.
\end{proposition}

A type D space-time admits two double null Debever directions
$\ell$, $k$. For any observer lying on the Weyl principal plane
$\{\ell,k\}$, the electric and magnetic Weyl tensors simultaneously
diagonalize. On the other hand, a type I space-time only admits an
observer with this property. Then, we have the following significant
and known result \cite{bel-3}:

\begin{proposition} \label{prop-or-g}
The observers at rest with respect to the gravitational field a
those for which the electric and magnetic Weyl tensors
simultaneously diagonalize.

In a type I space-time a unique observer $e_0$ at rest with respect
the gravitational field exists.

In a type D space-time the observers $e_0$ at rest with respect the
gravitational field a those lying on the Weyl principal plane.
\end{proposition}

The proper energy density defined in the electromagnetic case has
two qualities: firstly, it is the energy density measured by
observers at rest with respect the field and, secondly, it is the
minimum value of the relative energy density. Now, in the
gravitational case, we are taking one of these properties as a
definition and we will prove the other one as a theorem.

For a type I Bel-Robinson tensor $T$, the super-energy density
relative to the observer $e_0$ at rest with the field can be
obtained from expressions (\ref{T-xi-Omega}) and (\ref{Phi-A}) given
in \ref{C-theor-main} by taking $u=e_0$. Then, we obtain:
\begin{equation} \label{pe-g}
\tau_0 \equiv T(e_0,e_0,e_0,e_0) = \xi  \, ,
\end{equation}
where $\xi$ is defined in (\ref{alpha-xi}). Moreover,
the above expression is also valid in a type D space-time for any
observer $e_0$ at rest with the field.

Note that the right term of expression (\ref{pe-g}) is a scalar
invariant which can be defined in any space-time (radiative or not).
Then, we give the following definition:

\begin{definition} \label{def-pe-g}
We call {\em proper super-energy density} of a gravitational field
the invariant scalar
$$
\xi  \equiv \frac14 \sum t_i \, .
$$
\end{definition}

We know that for non radiative field, $\xi$ is the super-energy
density relative to an observer at rest with the field. Moreover,
from theorem \ref{theo-main} we obtain the following result.

\begin{theorem} \label{theo-min-g}
For a non radiative gravitational field (I or D) the minimum value
of the relative super-energy density is the proper super-energy
density $\xi$, which is acquired by the observers at rest with
respect to the field.
\end{theorem}

The meaning of the proper super-energy density $\xi$ for radiative
fields can be deduced by analyzing the asymptotic behavior of the
relative super-energy density. We can make this analysis using the
Bel-Robinson canonical forms presented in \ref{B-Bel-Robinson} and,
for type II, the expression (\ref{II-tau}). Then, we obtain:

\begin{theorem} \label{theo-inf-g}
For a radiative gravitational field (N, III or II) the relative
super-energy density tends to the proper super-energy density $\xi$
for observers traveling faster and faster in the direction
$\ell_{\bot}$, $\ell$ being the fundamental direction of the field.

For pure and asymptotic pure radiation (N or III), the proper
super-energy density $\xi$ is zero. For generic radiation (type II),
$\xi$ is strictly positive.
\end{theorem}

\ack We thank B Coll, A Garc\'{\i}a-Parrado, J A Morales-Lladosa and
J M M Senovilla for their comments. We are grateful to the referee
for his careful reading of the manuscript. His valuable suggestions
have improved the paper. This work has been supported by the Spanish
Ministerio de Ciencia e Innovaci\'on, MICIN-FEDER project
FIS2009-07705.

\appendix

\section{Notation}
\label{A-notation}

\begin{enumerate}
\item Composition of two 2-tensors $A,B$ as endomorphisms: $\ A \cdot B $, $ \
{(A \cdot B)^{\alpha}}_{\beta}= {A^{\alpha}}_{\mu}
{B^{\mu}}_{\beta}$.
\item In general, for arbitrary tensors $S,T$, $S \cdot T$ will be used to
indicate the contraction of adjacent indexes on the tensorial
product.
\item Square and trace of a 2-tensor $A$: $\ A^2 = A \cdot A,
\quad \tr A = {A^{\alpha}}_{\alpha}$.
\item The action on one or more vectors of an arbitrary tensor $S$ as
multilinear form will be denoted $S(x)$, $S(x,y)$, $S(x,y,z)$,...
For example, the action of a 2-tensor $A$ as an endomorphism $A(x)$
and as a bilinear form $A(x,y)$:
$$A(x)^{\alpha} ={A^{\alpha}}_{\beta} x^{\beta}, \qquad A(x,y) =
A_{\alpha \beta} x^{\alpha} y^{\beta} $$
\item The quadratic scalar associated with an arbitrary tensor $S$
will be denoted $(S,S)$. For example, if $x$ is a vector and $A$ is
a 2-tensor:
$$(x,x) = x^2 = g(x,x) \, , \qquad  (A,A) = A^{\alpha \beta}
A_{\alpha \beta}  \, . $$
\end{enumerate}

\section{Canonical forms of the Bel-Robinson tensor}
\label{B-Bel-Robinson}

The Bel-Robinson tensor $T$ defines an endomorphism on the space of
the traceless symmetric 2-tensors \cite{fsBR-1, fsBR-2}. The nine
eigenvalues $\{ t_k, \tau_k , \bar{\tau}_k \} $ depend on the three
(complex) Weyl eigenvalues $\{ \rho_k \}$ as $t_k = | \rho_k |^2$ ,
$\tau_k = \rho_i \bar{\rho}_j$, $(ijk)$ being a even permutation of
$(123)$.

Three independent invariant scalars can be associated with $T$ . In
fact, the nine eigenvalues $\{ t_i , \tau_{i} , \bar\tau_i \}$ can
be written in terms of three scalars $\{ p_i \}$ as \cite{fsBR-1}:
\begin{equation} \label{ti-pi}
t_i = - (p_j + p_k), \qquad \tau_i = p_i + \ci q, \qquad q^2=p_1 p_2 +
p_2 p_3 + p_3 p_1 \, .
\end{equation}
Note that scalars $\{ p_i \}$ satisfy the following inequalities:
\begin{equation} \label{pi->}
   p_j + p_k \leq 0, \quad  (j \neq k)  \, ; \qquad p_1 p_2 + p_2 p_3 + p_3
 p_1 \geq 0 \, .
 \end{equation}
Conversely, the scalars $p_i$ depend on the three real Bel-Robinson
eigenvalues $\{ t_i \}$ as:
\begin{equation} \label{pi-ti}
2 p_i= t_i - t_j - t_k , \qquad (i,j,k) \neq
\end{equation}

Studying the eigentensors of the Bel-Robinson tensor $T$ allows us
to obtain its canonical form for the different Petrov-Bel types
\cite{fsBR-2}. Now we summarize some of these results.

\subsection*{Type I}
\label{B-I}

Let $\{ e_{0} , e_{i} \}$ be the canonical frame of a type I Weyl
tensor. We can define the traceless symmetric 2-tensors:
\begin{equation}
\Pi_i = \frac{1}{2} ( v_i - h_i ), \qquad \Pi_{ij}= \frac{1}{2} (
e_{i} \stackrel{\sim}{\otimes} e_j + \ci \, e_{0}
\stackrel{\sim}{\otimes} e_k )   \, ,
\end{equation}
$(ijk)$ being a even permutation of $(123)$, and where $v_i = - e_0
\otimes e_0 + e_i \otimes e_i $, and $h_i = g - v_i$. Then,
$\{\Pi_{i}, \Pi_{jk}, \bar{\Pi}_{jk}\}$, is an orthonormal frame of
eigentensors of the Bel-Robinson tensor $T$. Moreover, $T$ takes the
canonical expression \cite{fsBR-2}:
\begin{equation} \label{can-I}
T= \sum_{i=1}^{3} t_i \, \Pi_{i} \otimes \Pi_{i} + \sum_{(ijk)}
\tau_i \, \Pi_{jk} \otimes \Pi_{jk} + \sum_{(ijk)} \bar\tau_i \,
\bar\Pi_{jk} \otimes \bar\Pi_{jk}  \, .
\end{equation}
where $(ijk)$ is a even permutation of $(123)$.

\subsection*{Type D}
\label{B-D}

Let $\{ e_{0} , e_{i} \}$ be a canonical frame of a type D Weyl
tensor, that is, the pairs $\{ e_{0} , e_{1} \}$ and $\{ e_{2} ,
e_{3} \}$ generate the Weyl principal 2-planes. Then, the
Bel-Robinson tensor $T$ takes the canonical expression (\ref{can-I})
with the eigenvalues restricted by $t_2 = t_3 = \tau_1 \not=0$, $t_1 = 4
t_2$ and $\tau_2 = \tau_3 = -2t_2$ \cite{fsBR-2}.

\subsection*{Type II}
\label{B-II}

A type II Weyl tensor admits a double null Debever direction $\ell$
and two simple ones $k_1, k_2$. Moreover, a time-like principal
plane exists which contains the direction $\ell$. Let $k$ be the
other null direction in the principal plane. Then, from an adapted
null frame of vectors $\{ \ell, k, m , \overline{m} \}$ we can
define the frame of 2-tensors  $\{ \Pi, \Lambda, {\rm K}, {\rm N},
\overline{{\rm N}}, \Omega, \overline{\Omega} , {\rm M},
\overline{{\rm M}} \}$ given by:
\begin{equation} \label{nfsta}
\hspace{-3mm}
\begin{array}{lll}
\Pi=- \frac{1}{2} ( \ell \stackrel{\sim}{\otimes}  k + m
\stackrel{\sim}{\otimes} \overline{m}), \quad  &  {\rm N}= -
\frac{1}{\sqrt{2}} \, \ell \stackrel{\sim}{\otimes}\overline{m},
\quad &
\Omega= \frac{1}{\sqrt{2}} \, k \stackrel{\sim}{\otimes} m , \\[1mm]
{\rm M}  = m \otimes m ,\quad &  \Lambda= - \, \ell \otimes \ell ,
\quad & {\rm K}= - \, k \otimes k  \, .
\end{array}
\end{equation}
Then, the Bel-Robinson eigenvalues are restricted by $t_2 = t_3 =
\tau_1\not=0$, $t_1 = 4 t_2$ and $\tau_2 = \tau_3 = -2t_2$, and the
Bel-Robinson tensor $T$ takes the canonical expression
\cite{fsBR-2}:
\begin{equation} \label{can-II}
\begin{array}{lcl}
T & = &  4 \,t_2  \, \Pi \otimes \Pi  - 2\, t_2 (
\Omega\stackrel{\sim}{\otimes} {\rm N} + \overline{\Omega}
\stackrel{\sim}{\otimes} \overline{{\rm N}} - {\rm N} \otimes {\rm
N} - \overline{{\rm N}} \otimes \overline{{\rm N}}
) \\[1mm] & + & t_2 (\Lambda \stackrel{\sim}{\otimes} {\rm K} + {\rm M}
\stackrel{\sim}{\otimes} \overline{{\rm M}} - \Lambda
\stackrel{\sim}{\otimes} \overline{{\rm M}} - \Lambda
\stackrel{\sim}{\otimes} {\rm M} + \Lambda \otimes \Lambda)  \, .
\end{array}
\end{equation}

\subsection*{Type III}
\label{B-III}

A type III Weyl tensor admits a triple null Debever direction $\ell$
and a simple one $k$. Then, from an adapted null frame of vectors
$\{ \ell, k, m , \overline{m} \}$ we can define the frame of
2-tensors $\{ \Pi, \Lambda, {\rm K}, {\rm N}, \overline{{\rm N}},
\Omega, \overline{\Omega} , {\rm M}, \overline{{\rm M}} \}$ given in
(\ref{nfsta}).
Then, all the eigenvalues vanish, and the Bel-Robinson tensor $T$
takes the canonical expression \cite{fsBR-2}:
\begin{equation} \label{BR-canIII}
T =   \Lambda \stackrel{\sim}{\otimes} \Pi + {\rm N}
\stackrel{\sim}{\otimes} \overline{{\rm N}} \, .
\end{equation}

\subsection*{Type N}
\label{B-N}

A type N Weyl tensor admits a quadruple null Debever direction
$\ell$. Then, all the eigenvalues vanish, and the Bel-Robinson
tensor $T$ takes the canonical expression \cite{fsBR-2}:
\begin{equation} \label{BR-canN}
T =  \ell \otimes \ell \otimes \ell \otimes \ell  \, .
\end{equation}

\section{Proof of theorem \ref{theo-main}}
\label{C-theor-main}

The proof of the theorem is based on the following two lemmas that
we will prove later.

\begin{lemma}  \label{lemma-II-III--N}
Let $T$ be a Bel-Robinson tensor of type II, III or N. Then, for any
observer $u$,
\begin{equation} \label{II-III--N}
T(u,u,u,u) > \xi = \frac12 \alpha   \, .
\end{equation}
\end{lemma}
\begin{lemma}  \label{lemma-I-D}
Let $T$ be a Bel-Robinson tensor of type I or D. Then, for any
observer $u$,
\begin{equation} \label{I-D}
T(u,u,u,u) \geq \xi \geq \frac12 \alpha   \, .
\end{equation}
\end{lemma}

\subsection*{Proof of theorem}

The first inequality in expression (\ref{T-alpha-xi}) of theorem
\ref{theo-main} is a direct consequence of the above two lemmas.

Now we are proving the second inequality in expression
(\ref{T-alpha-xi}) of theorem \ref{theo-main}. In \cite{fsBR-1} we
have introduced a second order superenergy tensor $T_{(2)}$
associated with the traceless part $W_{(2)}$ of the square $W^2$ of
the Weyl tensor $W$. That is, $T_{(2)}$ is defined as (\ref{BR-1})
by changing $W$ by $W_{(2)}$. It follows that $T_{(2)}$ has the same
properties as $T$ \cite{fsBR-1}. Then, we can apply to it the first
inequality in expression (\ref{T-alpha-xi}) already proved. Thus,
for any observer $u$:
\begin{equation} \label{T2-alpha}
T_{(2)}(u,u,u,u) \geq \frac14 \sqrt{(T_{(2)},T_{(2)})}  \, .
\end{equation}
From the specific expression of $T_{(2)}$ (see \cite{fsBR-1} for
more details) we can compute the left and the right terms of the
above inequality, and we obtain the second inequality in
(\ref{T-alpha-xi}).

\subsection*{Proof of lemma \ref{lemma-II-III--N}}
\label{C-rad}

The Bel-Robinson eigenvalues vanish for types N and III and,
consequently, both invariants $\xi$ and $\alpha$ vanish. Then,
inequality (\ref{II-III--N}) follows from the weak energy condition.

On the other hand, the eigenvalues and canonical form (\ref{can-II})
of a type II Bel-Robinson tensor lead to $2 \xi = 3 t_2 = \alpha$.
Moreover, an arbitrary observer $u$ can be written in the Weyl
canonical frame as $u = \lambda(e^{\phi} \ell + e^{-\phi} k) + \mu
(e^{\ci \sigma} m + e^{-\ci \sigma} \bar{m})$, $2(\lambda^2 - \mu^2)
= 1$. Then, using again (\ref{can-II}) we obtain,
\begin{equation} \label{II-tau}
\hspace{-2.3cm}T(u,u,u,u) = \xi ( 1 + 6 B)  , \quad B \equiv  2
\mu^2 + 4 \mu^4 \sin^2 2 \sigma + \left(\frac13 \lambda^2 e^{2 \phi}
\! - 2 \mu^2 \cos^2 2 \sigma \right)^2 \! \! > 0  .
\end{equation}
Thus, (\ref{II-III--N}) holds for type II space-times.

\subsection*{Proof of lemma \ref{lemma-I-D}}
\label{C-I-D}

Taking into account the eigenvalues relation (\ref{valprobel}) and
the Bel-Robinson canonical form (\ref{can-I}), we have
$$
16\, \xi^2\! = \! \Big(\!\sum_{i=1}^{3}\! t_i \Big)^2 \! \! =
\sum_{i=1}^{3}\! t_i^2 +2 \sum_{i < j}\! t_i t_j = \!
\sum_{i=1}^{3}\! t_i^2 + 2 \sum_{k=1}^{3}\! | \tau_k|^2 \geq \!
\sum_{i=1}^{3}[ t_i^2 +  \tau_i^2 +  \bar{\tau}_i^2] =(T,T)\! = 4
\alpha^2
$$
and then the second inequality in (\ref{I-D}) holds.

In order to prove the first one, let us write the observer $u$ in a
Weyl canonical frame $\{ e_{0} , e_{i} \}$, $u = u^{\alpha}
e_{\alpha}$, and let us calculate the relative super-energy density
by using the Bel-Robinson canonical form (\ref{can-I}). Then, we
obtain:
\begin{equation} \label{T-xi-Omega}
T(u,u,u,u) = \xi + \Omega \, , \qquad \Omega \equiv \Phi + 6 q A \,
,
\end{equation}
where $q$ is given in (\ref{ti-pi}) and
\begin{equation} \label{Phi-A}
\hspace{-2.3cm} \Phi \equiv A^i r_i , \quad r_i \equiv -(4 p_i + p_j
+ p_k) , \quad A^i \equiv (u^0)^2 (u^i )^2 - (u^j)^2 (u^k)^2 , \quad
A = \prod u^\alpha  ,
\end{equation}
$(i,j,k)$ being a even permutation of $(1,2,3)$. Thus, we must show
that $\Omega \geq 0$, an inequality which is a consequence of the
following two conditions:
\begin{equation} \label{Phi-R}
\Phi \geq 0 \, ,  \qquad  R \equiv \Phi^2 - (6 q A)^2 \geq 0 \, .
\end{equation}

Let us study the first one. From (\ref{Phi-A}) we obtain:
\begin{equation} \label{Phi-2}
\hspace{-0.8cm} \Phi  = - P^i p_i , \quad P^i \equiv  4 (u^0)^2
(u^i)^2 + (u^j)^2 + (u^k)^2 + ((u^j)^2 - (u^k)^2)^2 \geq 0 \, ,
\end{equation}
$(i,j,k)$ being a even permutation of $(1,2,3)$. At least one $t_i$
(says $t_3$) does not vanish in spacetimes of type I and D. Then,
from (\ref{pi->}) we have:
$$
- p_3 \geq  \frac{  p_1 p_2}{p_1 + p_2} \, ,
$$
and substituting in expression (\ref{Phi-2}) of $\Phi$ we obtain
$\displaystyle \Phi \geq - P^1 p_1 - P^2 p_2 + P^3 \frac{p_1
p_2}{p_1 + p_2}$, and then:
\begin{equation} \label{Phi-3}
\hspace{-1.9cm} \Phi  \geq - \frac{1}{p_1 + p_2} P^{AB} p_A p_B ,
\quad P^{AA} \equiv P^A , \ \ 2P^{12} \equiv P^1 + P^2 - P^3 , \quad
\texttt{\sc a,b} =1,2 \, .
\end{equation}
The quadratic form $P \equiv P^{AB} p_A p_B$ has principal minors of
order one which are non negative, $P^A \geq 0$ (see (\ref{Phi-2})),
and a straightforward calculation leads to a determinant
$\Delta_{P}$ which is also non negative,
$$\Delta_{P}\ = \ 9\left[ (u^0)^2 \sum_{i<j} ( u^i \,   u^j  )^2  +
\left[ 4 \sum_{i=1}^3 (u^i)^2 + 3 \right] \  \prod_{j=1}^3
(u^j)^2\right]\, \geq \, 0 \, .$$
Now we can apply the following theorem (see \cite{gantmacher}, page
309): {\it A quadratic form is non-negative if, and only if, all the
principal minors are non-negative}. Consequently we have $P \geq 0$,
and from (\ref{Phi-3}) and (\ref{pi->}), we obtain $\Phi \geq 0$.

Let us study now the second condition in (\ref{Phi-R}). Developing
the expression of $R$ we arrive to the quadratic form:
\begin{equation} \label{R}
\hspace{-2.0cm} R  =  R^{ij} r_i r_j , \quad R^{ii} \equiv [(u^0)^2
(u^i )^2 + (u^j)^2 (u^k)^2]^2 , \ \ R^{ij} \equiv A^i A^j - 4 A^2 ,
\ \ i,j=1,2,3 \, .
\end{equation}
where, in the expression of $R^{ii}$, $(i,j,k)$ is a even
permutation of $(1,2,3)$. Note that the principal minors of order
one are non negative, $R^{ii} \geq 0$. The principal minors of order
two $\Delta^{ij}$, and the determinant $\Delta$, are also non
negative:
$$
\begin{array}{lll}
\displaystyle \Delta^{ij} & = &  4 \, A^2   \left[ 1 + \displaystyle
(u^i)^2 + (u^j)^2 \right]^2 \,  \displaystyle \left[ ( u^i)^2 +
(u^j)^2
\right]^2 \, \geq \, 0 \, , \\[1mm]
\displaystyle \Delta & = & 64\,  A^4  \displaystyle \left[
\sum_{(ijk)} (u^i)^4 \displaystyle \left[ (u^j)^2 + (u^k)^2 \right]
+ \sum_{i<j} ( u^i \, \displaystyle  u^j  )^2 + 2 \, \prod_{i=1}^3
(u^i)^2 \right] \, \geq \, 0 \, .
\end{array}
$$
Consequently, applying again the stated theorem on non-negative
quadratic forms, we obtain $R \geq 0$.

\end{document}